\documentclass[preprint, 12pt]{elsarticle}

\usepackage{bm, color, amssymb, lineno, subfigure,amsmath}

\begin{document}

\begin{frontmatter}
\title{Steady state detection for computational fluid dynamics}

\author[hzdr,tuc]{Martin Boesler}
\author[hzdr]{Norbert Weber}
\ead{norbert.weber@hzdr.de}

\address[hzdr]{Helmholtz-Zentrum Dresden -- Rossendorf, %
Bautzner Landstr.\ 400, 
01328 Dresden, %
Germany}
\address[tuc]{Technische Universit\"at Chemnitz, Stra\ss e der Nationen 62,
  09111 Chemnitz, Germany}

\begin{abstract}
Large parameter studies of fluid dynamic instabilities can crucially
be simplified, if the user does not need to specify the simulation 
time. For this purpose, a steady state detection is implemented in 
the CFD library OpenFOAM. It terminates simulations automatically if
characteristic measurements do not change over time, i.e. when
saturation is reached. For that purpose, recent data is compared to
previous one using two selectable methods. Both calculate a value used
as steady state indicator. The first method performing a two sample
Student's t-test examines the difference between the means. Similarly,
the second method utilises a two sample f-test checking for changes
between the variances. Both methods are briefly described, compared
and their specific area of application is discussed. Their usefulness
is demonstrated with a simple exemplary test case.
\end{abstract}


\end{frontmatter}
\clearpage


\section{Introduction}
An instability arises when exceeding a certain critical threshold. It
grows then exponentially until reaching a saturated state. In order to
determine the characteristic properties of a fluid dynamic instability
(as, e.g., growth rate, steady state velocity) \emph{transient}
numerical simulation is required. A priori, the time for reaching the
saturated state is unknown. A steady state detection (SSD) or steady
state identification system can determine when saturation is reached
and switch off the simulation automatically. This saves space on the
hard drive. More importantly, an SSD system can drastically reduce user
interaction, because the simulation time does not need to be provided
any more. That will be especially useful when doing large parameter
studies, where the time to reach saturation depends on the
parameter. An SSD is not only useful for simulating instabilities, 
but for all transient flow phenomena.

A reasonable SSD method must satisfy several requirements: firstly, it
should not need any or very little user input. Secondly, it must be
robust, i.e. the saturated state must be detected securely. Thirdly,
it should be universal, i.e. different types of saturated states need
be detected. Finally, the SSD should be computationally simple and
easy to implement \cite{Padmanabhan2005,Rhinehart2013}.

Steady state detection is well established in process control. There,
deviations from the steady state, i.e. from the steady process need to
be detected. All common SSD methods use statistical approaches; the
most popular is the one by Cao and Rhinehart \cite{Cao1995}. For a short
introduction to different methods, see \cite{Bhat2004,Mhamdi2000} and for
a detailed overview \cite{Rhinehart2013}.

All common methods analyse a moving window \cite{Kim2008} of recent measurement
data. It does not exist an universal rule for the length of this
window \cite{Cao1995}: a long one will lead to a delayed steady state
detection, a short one will increase noise \cite{Bhat2004}. Applied to
numerical simulation, the ``measurement'' data may be, e.g., the volume
averaged velocity of each time step.

One simple way for SSD is performing a linear regression of the
measurement data inside the window. If steady, the slope should
ideally be zero. This can be verified easily with a t-test
\cite{Cao1995,Rhinehart2013}. Such a method is used by \cite{Wu2016}
while \cite{LeRoux2008} uses the first derivative of a polynomial
regression.

Another option is comparing the mean values of measurement data between
two subsequent windows. If the means do not change, the steady state
is reached. This hypothesis can easily be checked using a two sample
t-test \cite{Cao1995,Rhinehart2013} and is used e.g., by
\cite{Kelly2013}.

Alternatively, the standard deviations between two subsequent windows
may be compared using an f-test. In a steady state, the relation of
both should be one. A very similar approach consists in computing the standard
deviation in one window only, but using two different ways of
calculation. The hypothesis of a steady state is tested by an f-like
statistics, the r-test \cite{Cao1995}. For critical values for type I
and II errors see \cite{Cao1997,Shrowti2010}. This is the most
commonly used SSD method in process control
\cite{Yao2009,Bhat2004,Brown2000,Huang2014}.

Finally, also a wavelet approach may be used for SSD
\cite{Jiang2003,Mhamdi2000}. However, it is rather complex, and therefore
not discussed in detail here.

\section{Comparison of different SSD mechanisms}
Steady state detection in numerical simulation is different to process
control: typically, the ``measurement'' error in simulation is very
small compared to reality. While process control aims in detecting a
deviation of the steady process, we search the steady state itself.

A saturated flow can generally be classified into three different
categories: firstly, it may be really steady, with ideally no
fluctuations. Secondly, it may be periodically oscillating and thirdly
turbulent. Detecting these different types of saturated states with one
method will be challenging. Therefore, we focus in this article
as a first step on real steady states only.

Table \ref{t:comparison} illustrates how different tests fulfil the
requirements to SSD described in the introduction. A linear regression
test will work generally fine except for oscillating or turbulent
flow. However, it is computationally expensive -- a disadvantage
which it has in common with the wavelet test. Anyway, the latter might
be interesting for detecting unsteady saturated flows.

The t-test is
robust, simple and universal. It may even be used for unsteady
saturated flows. However, its application to oscillating flows is
limited: the moving window must be large enough to include a full
period. Further, an oscillating flow with changing amplitude might be
detected as steady, as long as the mean does not change. This drawback
can be compensated using an f-test. It will detect falling or rising
oscillation periods. However, as it compares only standard deviations
it can not detect changing means.

In many cases, a double t-test will be the optimal method for steady
state detection. A composite f-t-test, which compares both, means and
standard deviations, will be even safer in the sense of avoiding a
false steady state detection.

\begin{table}[t!]\vspace{8pt}
\centering
\caption{Requirements to steady state detection and performance of
  different SSD methods (+ means good, -- bad).} 
\begin{tabular}{lcccc}\hline\label{t:comparison}
criteria&linear regression&t-test&f-test&wavelet\\
\hline
  robustness &?&+&--&?\\
  universality &--&+&+&--\\
  low user interaction &?&?&?&?\\
  computationally simple &--&+&+&--\\
\hline\end{tabular}\vspace{3pt}\end{table}

\section{Theory and implementation}
We will describe the double t-, f- and f-t-composite test in the
following. All of them check the null hypothesis of a steady state.
In a first step, the size of two consecutive moving windows needs to
be defined. Often, the time step of the simulation is determined by
the flow velocity using some kind of Courant Friedrich Levy number. It
is therefore reasonable to use a fixed number of $n$ time steps to
define the size of each moving window. The arithmetic mean of a
property $x$ (e.g. volumetric averaged velocity) of the first and
second window is defined as
\begin{equation}
  \overline x_1 = \frac{1}{n}\sum_{i=k-2n+1}^{k-n} x_i\qquad\text{and}\qquad
  \overline x_2 =
  \frac{1}{n}\sum_{i=k-n+1}^{k} x_i
\end{equation}
with $k$ denoting the current time step index. Similarly, the variances are
\begin{align}
  s_1^2&=\frac{1}{n-1}\sum_{i=k-2n+1}^{k-n}\left(x_i - \overline x_1\right)^2\quad\text{and}\\
  s_2^2&=\frac{1}{n-1}\sum_{i=k-n+1}^{k}\left(x_i - \overline x_2\right)^2.
\end{align}
After the simulation has run through the first $2n$ time steps, the
means and variances are computed using the old values of $\overline x$ and
$s$ and the new data point $x_k$ as
\begin{align}
  \overline x_{1,k}&=\overline x_{1,k-1}-\frac{x_{k-2n}-x_{k-n}}{n}\\[0.5em]
  \overline x_{2,k}&=\overline x_{2,k-1}-\frac{x_{k-n}-x_{k}}{n}\\[0.5em] 
  s_{1,k}^2&=s_{1,k-1}^2-\frac{x_{k-2n}^2-x_{k-n}^2-n\cdot(\overline
             x_{1,k-1}^2-\overline x_{1,k}^2)}{n-1}\\[0.5em] 
  s_{2,k}^2&=s_{2,k-1}^2-\frac{x_{k-n}^2-x_{k}^2-n\cdot(\overline
             x_{2,k-1}^2-\overline x_{2,k}^2)}{n-1}.
\end{align}
Please note that the point $x_{k-2n}$ is just outside of both windows.
This procedure is more efficient than recalculating means and
variances from all time steps \cite{Kim2008}. 

The t-test compares the means between two consecutive moving
windows. In order to obtain a dimensionless value, the mean is divided
by the square root of the variances. The test statistics is
\begin{equation}
  T = \frac{|\overline x_2-\overline x_1|}{\sqrt{s_1^2 + s_2^2}}\sqrt{n}.
\end{equation}
For a perfectly steady state, $T$ will be zero. However, the variances
will be zero, too. In that case, $T$ will be infinite. To avoid such
kind of problems, we add an artificial error of 1\,\% to the
data points when calculating the variances.

The f-test compares the variances between the two sampling
windows. The test statistics is defined as
\begin{equation}
  F =
  \frac{\text{max}(s_1^2,s_2^2)}{\text{min}(s_1^2,s_2^2)}. 
\end{equation}
At steady state, $F$ will ideally be one. Again, $\sigma^2$ should
not get zero and therefore a 1\,\% artificial error is added during it's
calculation.

The combined f-t test checks first the f- and thereafter the
t-test. Only if both detect a steady state, the simulation
ends. Critical values for $T$ and $F$ can be given by statistical
approaches \cite{Cao1996,Shrowti2010} or may be chosen empirically
after doing a single test simulation as described in the following
section. Further, an appropriate number of samples ($n$) must be
selected. 

\section{OpenFOAM Testcase}
We use the Tayler instability (TI) \cite{Vandakurov1972} as exemplary test case
for the developed steady state detection system. When an electrical
current flows through a liquid conductor, the resulting Lorentz force
is always directed to the centre of the conductor. This configuration
is unstable if a certain critical current is exceeded. The TI is not
only discussed in context of a possible explanation of the 11-years
solar cycle
\cite{Bonanno2012,Seilmayer2012,Weber2015b,Stefani2016,Stefani2017}
but also as some upper limit for the design of liquid metal batteries
\cite{Stefani2011,Herreman2015,Weier2017}. The latter are proposed as
a cheap grid-scale energy storage for fluctuating renewable energies
\cite{Kim2013b}. 

\begin{figure}[bth]
\centering
\subfigure[]{\includegraphics[height=5cm]{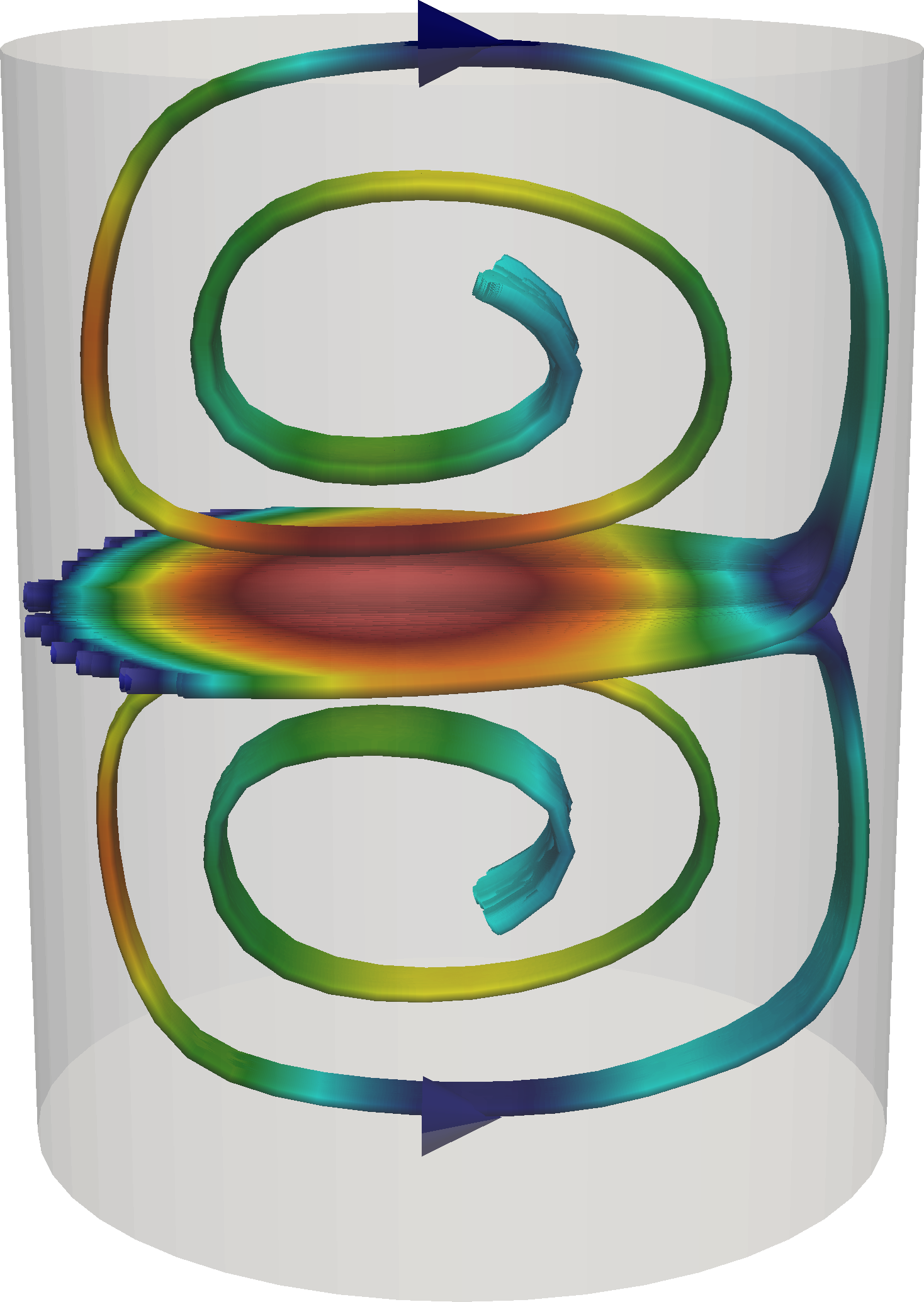}}\hfill
\subfigure[]{\includegraphics[height=5cm]{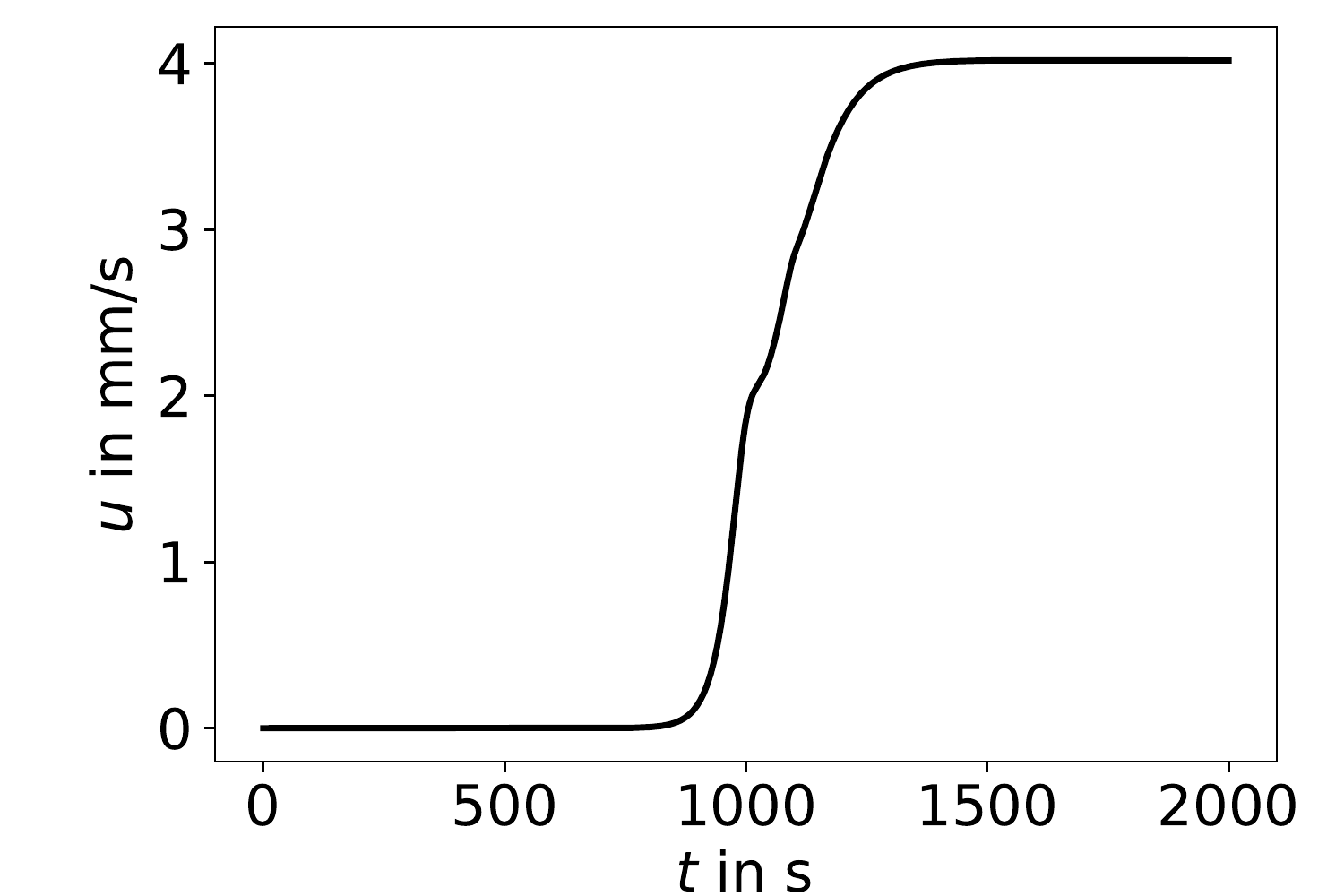}}\vspace{3mm}
\subfigure[]{\includegraphics[width=0.5\textwidth]{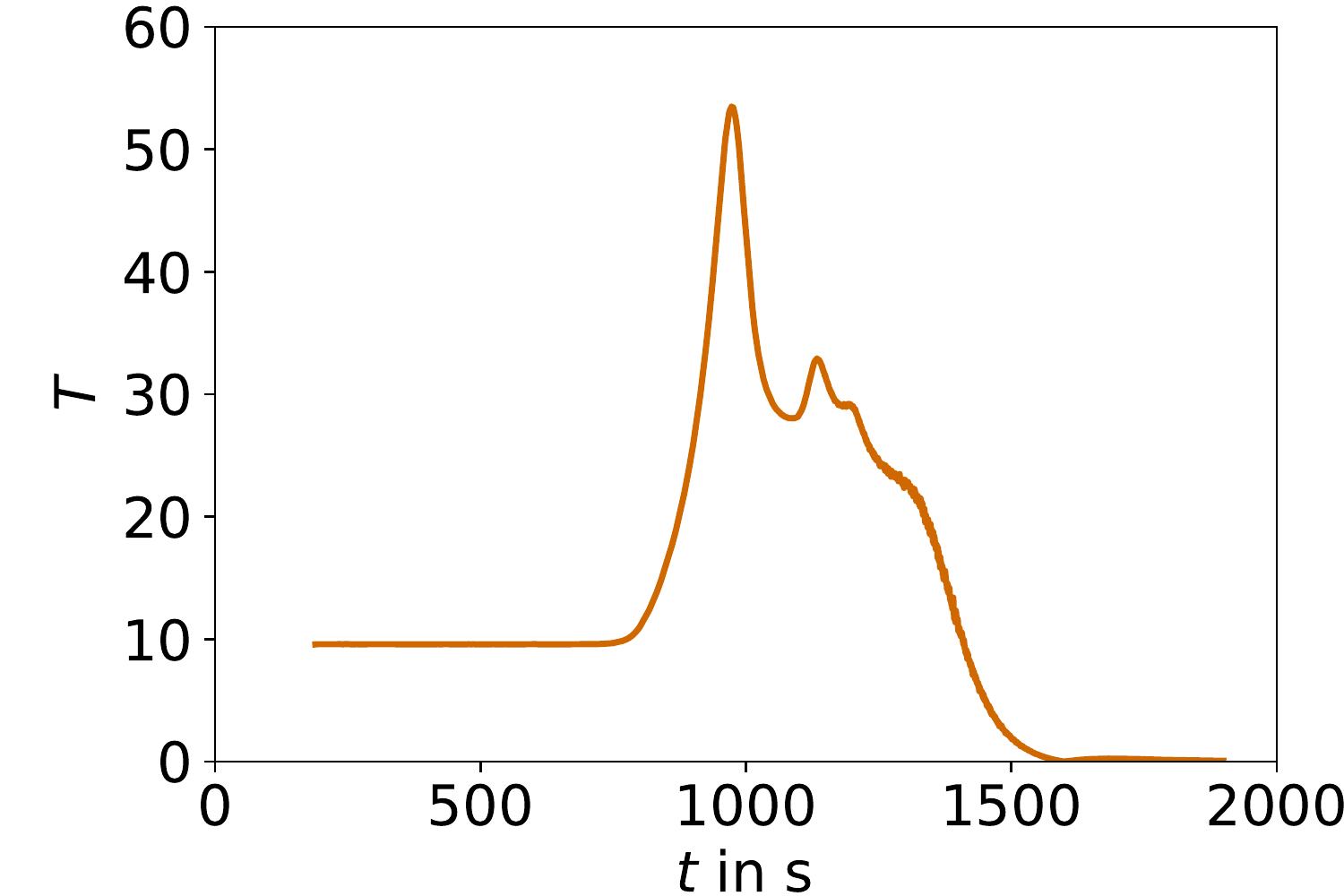}}\hfill
\subfigure[]{\includegraphics[width=0.5\textwidth]{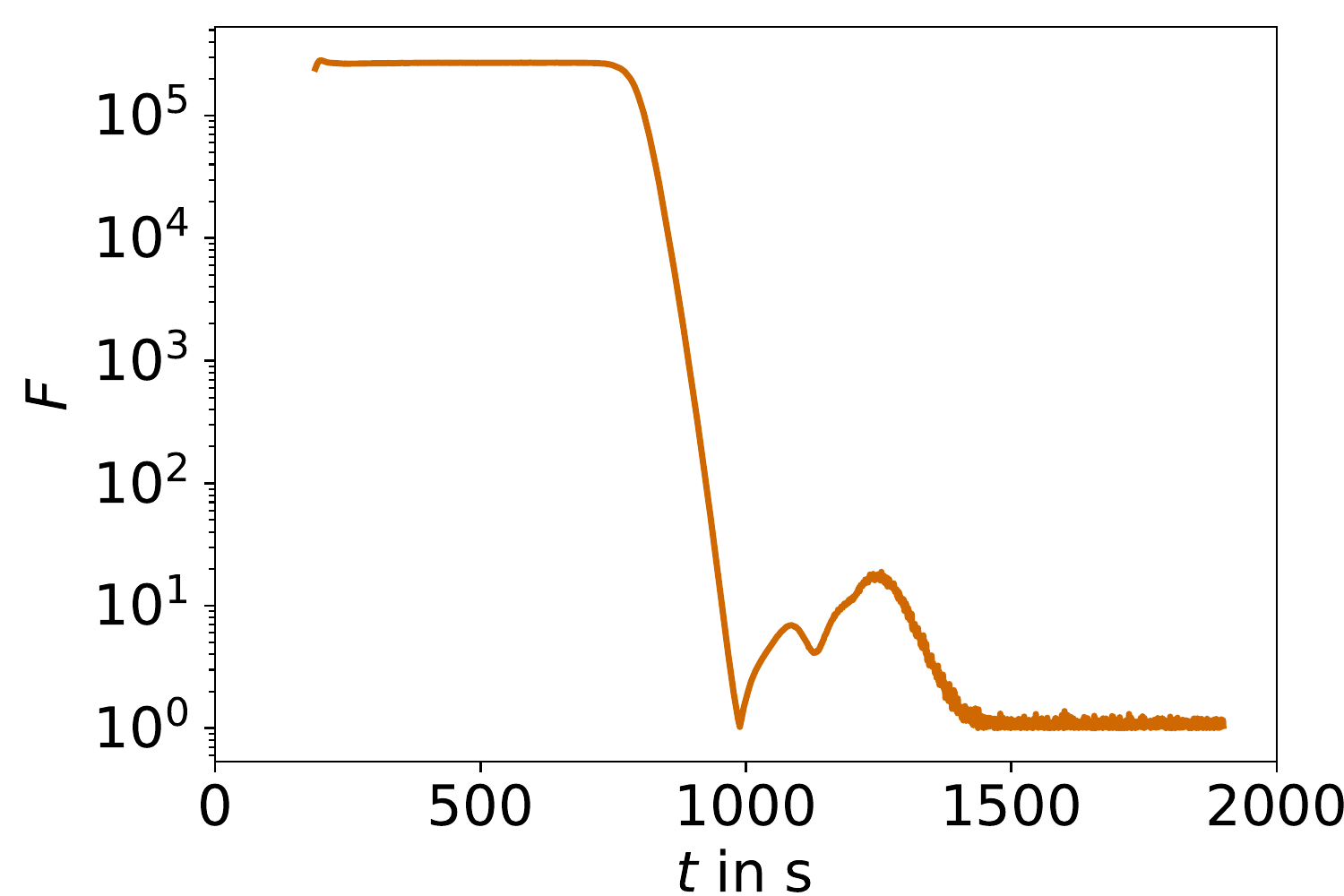}}
\caption{General flow structure of the Tayler instability in a
  cylindrical vessel (a), volume averages velocity over time for
  $I=4$\,kA (b) and test statistics over time for the t- (c) and f-test
(d).}
\label{f1}
\end{figure}

We simulate the TI in a cylindrical liquid metal conductor of diameter
$d=0.1$\,m, height $h=0.125$\,m, density $\rho = 600\,$kg/m$^3$,
kinematic viscosity $\nu = 2.5\cdot10^{-7}$\,m$^2$/s and electrical conductivity
$\sigma = 3.2\cdot10^6$\,S/m. Figure \ref{f1}a shows the general flow
structure and figure \ref{f1}b the volumetric averaged velocity over
time for an applied electric current of $I=4$\,kA. The exponential
growth phase and the steady state can be observed very well. We apply
the statistical tests using the volume averaged velocity.

It can be well observed in fig. \ref{f1}c that the test statistics of
the t-test goes to a value close to zero only in the saturated
state. Similarly, the test statistics of the f-test (fig. \ref{f1}d)
reaches a value close to one. However, the latter is less reliable, as
it detects already a false steady state during the growth of the
instability.

Many test runs revealed that $F$ depends only slightly and $T$ not at
all on the applied current of the Tayler instability. However, both
statistics depend strongly on the number of samples $n$, i.e. on the
size of the moving window. While the t-test works already well with
only ten sampling values, the f-test needs at least 100
samples. Such a large number of samples allows for a better
distinction between a growing and steady flow. Unfortunately, also the
steady state value of the statistics $T$ and $F$ depends on the number
of samples. For the particular case of the Tayler instability we
recommend $F_\text{cr}=1.6$ with $n=100$ and $T_\text{cr}=n\cdot5\cdot
10^{-4}$ with $n\ge 10$. A steady state is likely reached, when $T$
and $F$ fall below of $T_\text{cr}$ and $F_\text{cr}$. The probably best way of adapting the SSD to
other instabilities is doing a test simulation with a coarse
grid. Reasonable critical values for $T$ and $F$ can easily be
determined using a simple python script.

\section{Summary and outlook}
We have discussed where and why a steady state detection (SSD) system
can be beneficial in computational fluid dynamics. We have further
specified the requirements for a suitable method. After categorising
saturated flow states into steady, oscillating and turbulent we have
focused our attention to purely steady states, only. We further
compared different statistical SSD methods and selected a double
f- and t-test as the most promising approaches. The tests compare the
means and variances of two consecutive moving windows of simulation data. A
steady state is detected when they do not change any more. We have
further demonstrated the usefulness of the developed SSD system using
the Tayler instability as exemplary case. We have shown, that the
proper selection of the number of sample points $n$ and the critical
value of the test-statistics $F_\text{cr}$ and $T\text{cr}$ is the
most challenging part. We have provided typical values of all three
variables. 

In a next step, general rules for selecting $n$, $F_\text{cr}$ and
$T_\text{cr}$ shall be developed. The SSD library will be extensively
tested and continuously improved\footnote{A current version with a well
  documented source code can be obtained from the authors.}. In a
first step it is planned to use it for other instabilities, as, e.g.,
Rayleigh-B\'enard convection \cite{Shen2015,Koellner2017} and the
metal pad roll instability in liquid metal batteries
\cite{Zikanov2015,Weber2017,Horstmann2017}. On the long run, the SSD
system is planned to be extended to cope with oscillating and
turbulent saturated flows.

\section*{Acknowledgements}
This work was supported by Helmholtz-Gemeinschaft Deutscher
For\-schungs\-zentren (HGF) in frame of the Helmholtz Alliance
``Liquid metal technologies'' (LIMTECH).  Fruitful discussions with
V. Galindo, T. Gundrum and T. Weier on several aspects of
instabilities and steady state detection are gratefully acknowledged. 

\section*{References}
\bibliographystyle{elsarticle-num}
\bibliography{literature}

\end{document}